\documentclass[a4paper,aps,prd,nofootinbib,onecolumn,notitlepage]{revtex4}
\RequirePackage[english]{babel}
\RequirePackage[latin1]{inputenc}
\RequirePackage[T1]{fontenc}
\RequirePackage{mathrsfs}
\RequirePackage{amsmath}
\RequirePackage{amssymb}
\RequirePackage{amsbsy}
\RequirePackage{bm}
\usepackage[lofdepth,lotdepth]{subfig}
\usepackage{graphicx}
\usepackage{multirow} 
\usepackage{caption}
\def\de#1/de#2{\frac{\partial {#1}}{\partial {#2}}}
\def\a{\alpha}
\def\b{\beta}

\def\m{\mu}
\def\n{\nu}

\def\d{\delta}

\newcommand{\nn}{\nonumber}
\newcommand{\ba}{\begin{eqnarray}}
\newcommand{\ea}{\end{eqnarray}}
\newcommand{\be}{\begin{equation}}
\newcommand{\ee}{\end{equation}}

\begin{document}
\title{Anisotropic fluid spheres in Ho\v rava gravity and Einstein-\ae ther theory with a non-static \ae ther}
\author{Daniele Vernieri}
\affiliation{Instituto de Astrof\'isica e Ci\^encias do Espa\c{c}o, 
Faculdade de Ci\^encias da Universidade de Lisboa, Campo Grande, PT1749-016 Lisboa, Portugal}
\date{\today}
\begin{abstract}
In this paper we consider spherically symmetric interior spacetimes filled by anisotropic fluids in the context of Ho\v rava gravity and Einstein-\ae ther theory. We assume a specific non-static configuration of the \ae ther vector field and show that the field equations admit a family of exact analytical solutions which can be obtained if one of the two metric coefficients is assigned. We study as an illustrative example the case in which the metric of the interior spacetime reproduces the Newtonian potential of a fluid sphere with constant density. 
\end{abstract}
\maketitle


\section{Introduction} 

Ho\v rava gravity was proposed in 2009 as a power-counting renormalizable theory of quantum gravity~\cite{Horava:2009uw,Blas:2009qj}. In the past years much work has been done to show that the theory is renormalizable~\cite{DOdorico:2014tyh,Barvinsky:2015kil,Barvinsky:2017zlx,Barvinsky:2019rwn} beyond the power-counting arguments~\cite{Visser:2009fg,Sotiriou:2009gy,Sotiriou:2009bx,Vernieri:2011aa,Vernieri:2012ms,Vernieri:2015uma}. Ho\v rava gravity has also been severely constrained by means of some tests at both astrophysical~\cite{Yagi:2013qpa,Yagi:2013ava,Ramos:2018oku} and cosmological scales~\cite{Audren:2014hza,Frusciante:2015maa}, and it passes all of them with flying colors. Moreover it is also consistent with the constraint on the speed of propagation of gravitational waves coming from the near-simultaneous temporal and spatial observation of the gravitational-wave event GW170817 and the gamma-ray burst GRB 170817A~\cite{Monitor:2017mdv,Gumrukcuoglu:2017ijh}.
The theory breaks Lorentz invariance at any energy scale since a preferred direction is naturally encoded in its formulation. This locally amounts to having a timelike hypersurface-orthogonal \ae ther vector field which is defined in each point of the spacetime. If one considers the low-energy limit of Ho\v rava gravity in a covariant form, the latter proves to be equivalent to Einstein-\ae ther theory~\cite{Jacobson:2000xp} once the \ae ther vector is taken to be hypersurface-orthogonal at the level of the action~\cite{Jacobson:2010mx}. In spherical symmetry, any vector is automatically hypersurface-orthogonal; therefore by virtue of this fact it can be shown that the two theories share the same solutions in such a background~\cite{Blas:2010hb}. Because of the intrinsic highly non-linear structure of the field equations, only a few analytical and numerical solutions are known both in vacuum~\cite{Barausse:2011pu,Berglund:2012bu,Barausse:2012qh,Wang:2012nv,Barausse:2013nwa,Sotiriou:2014gna} and inside matter~\cite{Eling:2006df,Eling:2007xh,Vernieri:2017dvi,Vernieri:2018sxd}. Thus, it is really necessary to focus more effort in this direction, since many of the phenomenological implications of the theory are still unknown, even in highly symmetric spacetimes.
For this purpose in the present manuscript we consider spherically symmetric interior spacetimes filled by anisotropic fluids~\cite{Herrera:1997plx,Harko:2002db,Carloni:2017bck} in the context of the low-energy limit of Ho\v rava gravity.
The approach that we undertake here is similar to the one used in Refs.~\cite{Vernieri:2017dvi,Vernieri:2018sxd} in which the equation of state of the inner fluid is left unspecified, but after a viable solution has been found, it can be instead reconstructed {\it a posteriori} by using the same method exploited in Ref.~\cite{Vernieri:2018sxd}. This approach generically looks more realistic since, despite all the work that has been done until now, we still lack a proper modeling of the interior spacetime of relativistic objects (see Ref.~\cite{Ozel:2016oaf} and references therein). Then it seems appropriate to leave unspecified the equation of state relating the thermodynamical quantities, whose study is anyhow out of the scope of the present paper.
In Refs.~\cite{Vernieri:2017dvi,Vernieri:2018sxd}, this kind of setting has already been studied, but in the more restricted case of a static \ae ther, which means that the \ae ther vector is aligned with the timelike Killing vector and then has only one non-vanishing component. Here we consider instead a more general ansatz where the \ae ther vector field indeed has two non-trivial components. We derive the corresponding field equations and find a strategy to analytically solve them by means of some choice of the \ae ther components and an appropriate redefinition of variables. In this framework, we show that the field equations admit a family of infinite exact analytical solutions, characterized by choosing arbitrarily one of the metric coefficients of the spherically symmetric interior spacetime.

In Sec.~\ref{Sec1}, we present the gravitational action and the field equations of Ho\v rava gravity and discuss the equivalence of its low-energy limit to Einstein-\ae ther theory when the \ae ther is chosen to be hypersurface-orthogonal at the level of the action. In Sec.~\ref{Sec2}, we introduce the spherically symmetric background metric, the corresponding \ae ther vector field, and the stress-energy tensor suitable for the anisotropic fluid description. In Sec.~\ref{Sec3}, we write down explicitly the system of the independent field equations in terms of the metric, the \ae ther components, and the thermodynamical quantities. In Sec.~\ref{Sec4} we focus on a specific case in which the interior metric reproduces the Newtonian potential of a fluid sphere with constant density. In Sec.~\ref{Sec5}, the conclusions are drawn.

\section{Ho\v rava gravity and Einstein-\ae ther theory}
\label{Sec1}

The action of Ho\v rava gravity~\cite{Horava:2009uw,Blas:2009qj} as written in the preferred foliation looks like
\be \label{horava}
\mathcal{S}_{H}=\frac{1}{16\pi G_H}\int{dT d^3x\sqrt{-g}\left(K_{ij}K^{ij}-\lambda K^2 +\xi \mathcal{R}+\eta a_i a^i+\frac{L_4}{M_\ast^2}+\frac{L_6}{M_\ast^4}\right)}+S_m[g_{\mu\nu},\psi],
\ee 
where $G_H$ is the effective gravitational coupling constant; $g$ is the determinant of the metric $g_{\mu\nu}$; $\mathcal{R}$ is the Ricci scalar of the three-dimensional constant-$T$ hypersurfaces; $K_{ij}$ is the extrinsic curvature and $K$ is its trace; and $a_i=\partial_i \mbox{ln} N$, where $N$ is the lapse function and $S_m$ is the matter action for the matter fields collectively denoted by $\psi$. The couplings $\left\{\lambda,\xi,\eta\right\}$ are dimensionless, and general relativity (GR) is identically recovered when they take the values $\left\{1,1,0\right\}$, respectively. Finally, $L_4$ and $L_6$ collectively denote the fourth-order and sixth-order operators that make the theory power-counting renormalizable, and $M_\ast$ is a characteristic mass scale which suppresses them. 

In what follows, we consider the covariantized version of the low-energy limit of Ho\v rava gravity, named the {\it khronometric} model, that is obtained by discarding the higher-order operators in $L_4$ and $L_6$. In order to write it in a covariant form, let us consider the action of Einstein-\ae ther theory~\cite{Jacobson:2000xp}; that is,
\be
\mathcal{S}_{\mbox{\scriptsize\ae}}=\frac{1}{16\pi G_{\mbox{\footnotesize \ae}}}\int{d^4x\sqrt{-g}\left(-R+L_{\mbox{\scriptsize\ae}}\right)}+S_m[g_{\mu\nu},\psi], \label{aetheraction}
\ee
where $G_{\mbox{\footnotesize \ae}}$ is the ``bare'' gravitational constant; $R$ is the four-dimensional Ricci scalar; $u^a$ is a timelike vector field of unit norm, {\it i.e.}, $g_{\mu\nu}u^\mu u^\nu=1$, from now on referred to as the ``\ae ther''; and 
\be
L_{\mbox{\scriptsize\ae}}=-M^{\a\b}{}_{\m\n} \nabla_\a u^\m \nabla_\b u^\n\,,
\ee
with $M^{\a\b}{}_{\m\n}$ defined as
\be 
M^{\a\b}{}_{\m\n} = c_1 g^{\a\b}g_{\m\n}+c_2\d^{\a}_{\m}\d^{\b}_{\n}+c_3 \d^{\a}_{\n}\d^{\b}_{\m}+c_4 u^\a u^\b g_{\m\n}\,,
\ee
where $c_i$'s are dimensionless coupling constants.

Then, one can take the \ae ther vector to be hypersurface-orthogonal at the level of the action, which locally amounts to choosing
\be
u_\alpha=\frac{\partial_\alpha T}{\sqrt{g^{\mu\nu}\partial_\mu T \partial_\nu T}}\,,
\ee
where in the covariant formulation, the preferred time $T$ becomes a scalar field (the {\it khronon}) which defines the preferred foliation. Finally, the two actions in Eqs.~\eqref{horava} and~\eqref{aetheraction} are shown to be equivalent if the parameters of the two respective theories are mapped into each other as~\cite{Jacobson:2010mx}
\be
\label{eqn:corresp}
\frac{G_H}{G_{\scriptsize\mbox{\ae}}}=\xi = \frac{1}{1-c_{13}}\,, \hspace{2em} \frac{\lambda}{\xi} = 1 + c_2\,, \hspace{2em} \frac{\eta}{\xi} = c_{14}\,,
\ee
where $c_{ij} = c_i+c_j$. In what follows, we consider the covariant formulation of Ho\v rava gravity in order to perform the calculations. \\

The variation of the action in Eq.~\eqref{aetheraction} with respect to $g^{\a\b}$ and $T$ yields, respectively~\cite{Ramos:2018oku},

\ba 
\label{eqhorava}
&&G_{\a\b} - T^{\mbox{\footnotesize \ae}}_{\a\b}=8\pi G_{\mbox{\footnotesize \ae}} T^{m}_{\a\b}\,,\\
\label{hleq}
&&\partial_\mu \left(\frac{1}{\sqrt{\nabla^\alpha T \nabla_\alpha T}}\sqrt{-g} \mbox{\AE}^\mu \right)=0\,,
\ea
\vskip 0.2cm
where
$G_{\a\b}=R_{\a\b}-R g_{\a\b}/2$ is the Einstein tensor,
\ba \label{Tae}
T^{\mbox{\footnotesize \ae}}_{\a\b}&=&\nabla_\m\left(J^{\phantom{(\a}\m}_{(\a}u_{\b)}-J^\m_{\phantom{\m}(\a}u_{\b)}-J_{(\a\b)}u^\m\right)+c_1\,\left[ (\nabla_\m u_\a)(\nabla^\m u_\b)-(\nabla_\a u_\m)(\nabla_\b u^\m) \right]\nonumber\\
&&+\left[ u_\n(\nabla_\m J^{\m\n})-c_4 \dot{u}^2 \right] u_\a u_\b
+c_4 \dot{u}_\a \dot{u}_\b-\frac{1}{2} L_{\mbox{\scriptsize\ae}} g_{\a\b} + 2 \mbox{\AE}_{(\a}u_{\b)}
\ea
is the khronon stress-energy tensor, 

\be
J^\a_{\phantom{a}\m}=M^{\a\b}_{\phantom{ab}\m\n} \nabla_\b u^\n\,,
\qquad\dot{u}_\n=u^\m\nabla_\m u_\n\,,\qquad \mbox{\AE}_\mu = \left(\nabla_\a J^{\a\n}-c_4\dot{u}_\a\nabla^\n u^\a\right) \left(g_{\mu\nu}-u_\m u_\n\right)\,,
\ee
\vskip 0.3cm
and $T^{m}_{\a\b}$ is the matter stress-energy tensor, defined as
\be
T^{m}_{\a\b} = \frac{2}{\sqrt{-g}}\frac{\delta S_m}{\delta g^{\a\b}}\,.
\ee

\section{Spherically symmetric metric, anisotropic fluids, and a non-static \ae ther}
\label{Sec2}

In spherical symmetry, the most general metric can be written as
\be
ds^2 = A(r) dt^2 - B(r) dr^2 - r^2\,\big(d\theta^2+\sin^2\theta d\phi^2\big). \label{Eq0}
\ee
Moreover, in what follows, we will consider the interior spacetime of a fluid sphere filled by an anisotropic fluid whose stress-energy tensor is given by 

\be
T_{\mu\nu}=\left(\rho + p_t\right) v_\mu v_\nu - p_t g_{\mu\nu} + \left(p_r-p_t\right) s_\mu s_\nu\,,
\ee
\vskip 0.2cm
\noindent
where $\rho$ is the density, $p_r$ and $p_t$ are the radial and transversal pressure, respectively, $v^\mu$ denotes the 4-velocity of the fluid
\be
v^\mu =\biggl(\frac{1}{\sqrt{A(r)}},0,0,0\biggr),
\ee
and $s^\mu$ is a spacelike 4-vector defined as
\be
s^\mu = \biggl(0,\frac{1}{\sqrt{B(r)}},0,0\biggr),
\ee
\vskip 0.2cm
\noindent satisfying the relations $s^\mu s_\mu = -1$ and $s^\mu u_\mu = 0$.
It can be easily shown that the components of the stress-energy tensor can be explicitly written as
\be
T_\mu^{\phantom{\mu}\nu} = \mbox{diag}\big(\rho,-p_r,-p_t,-p_t\big).
\ee 
\vskip 0.1cm
\noindent The \ae ther vector field, which is by definition a timelike vector of unit norm, in spherical symmetry is always hypersurface-orthogonal and takes the following general form: 

\be
u^\alpha =\biggl(F(r),\sqrt{\frac{A(r) F(r)^2-1}{B(r)}},0,0\biggr). \label{aether}
\ee
\vskip 0.1cm
\noindent
The independent field equations that we have to consider are the modified Einstein equations (0-0), (1-1), (1-2) and (2-2) in Eq.~\eqref{eqhorava}; the equation for the scalar field $T$ in Eq.~\eqref{hleq}; and the conservation equation for the stress-energy tensor of anisotropic matter that is,
\be
p_r'(r)+\left[\rho(r)+p_r(r)\right]\frac{A'(r)}{2A(r)}=\frac{2}{r} \left[p_t(r)-p_r(r)\right]. \label{conserv}
\ee
\vskip 0.1cm
\noindent
The expressions of the field equations are very long and awful, so they are not displayed here. However, through a direct inspection it is quite straightforward to notice that the field equations are considerably simplified by making the choice $F(r)=\frac{q}{\sqrt{A(r)}}$, and the \ae ther vector in Eq.~\eqref{aether} then becomes
\be
u^\alpha =\biggl(\frac{q}{\sqrt{A(r)}},\sqrt{\frac{q^2-1}{B(r)}},0,0\biggr). \label{aether2}
\ee
Let us stress that, even by implementing this specific assumption, the \ae ther vector field has anyhow two non-trivial components. Then, it is still more general than the \ae ther vector widely used in literature and referred to as ``static \ae ther'' which is by definition aligned with the timelike Killing vector. The choice of a static \ae ther is just a special case which is included in the more general framework developed here by setting $q=1$ in Eq.~\eqref{aether2}.

\section{Field equations}
\label{Sec3}

In order to write down the field equations in a more compact form, let us consider the following redefinition of variables:
\be
Y(r)=r\frac{A'(r)}{A(r)}\,, \,\,\,\,\,\,\,\, W(r)=r\frac{B'(r)}{B(r)}\,.
\ee
\vskip 0.1cm
\noindent
Then, the (0-0) component of the modified Einstein equations in Eq.~\eqref{eqhorava} becomes

\ba
&&\frac{1}{8 \xi (\eta -\lambda +1) r^2 B(r)}\biggl[8 \xi  B(r) (\eta -\lambda +1)+8 \left(-\eta +2 \lambda ^2-3 \lambda +q^2 (\eta -\lambda +1) (2 \lambda -\xi -1)+1\right)\biggr. \nn \\
&&\biggl.+Y(r) \left(\lambda +q^2 (\eta -\lambda +1)-1\right) (8 (\lambda -\xi )+(-\eta +\lambda -1) Y(r))+8 W(r) (\eta  \lambda -\lambda  \xi +\xi )\biggr] = 8 \pi G_{\text{ae}} \rho(r)\,, \label{Y1}
\ea
the (1-1) component is

\ba
&&\frac{1}{8 \xi  r^2 B(r)}\biggl[-8 \xi  B(r)+16 \lambda +8 q^2 (-2 \lambda +\xi +1)+ \left(\lambda +q^2 (\eta -\lambda +1)-1\right)Y(r)^2\biggr. \nn \\
&&\biggl.+8 \left(\lambda -\lambda  q^2+\xi  q^2\right)Y(r)-8\biggr] = 8 \pi G_{\text{ae}} p_r(r)\,, \label{Y2}
\ea
and the (2-2) component is

\ba
&&\frac{1}{8 \xi (\eta -\lambda +1) r^2 B(r)}\biggl[16 (\lambda -1) \left(\lambda  \left(q^2-1\right)-\xi  q^2\right)+4 W(r) \left(-2 \eta  \lambda +\eta +2 \lambda  \xi -3 \lambda +q^2 \left(-\xi  (\eta +3 \lambda +1)+2 \eta  \lambda \right.\right.\biggr. \nn \\
&&\left.\left.-\eta +3 \lambda +2 \xi ^2-1\right)+1\right) +Y(r) \left(-4 \xi  \left(q^2 (\eta +3 \lambda +1)-2 \lambda \right)+4 \left(q^2-1\right) (\eta +\lambda  (2 \lambda -1)+1)+8 \xi ^2 q^2\right. \nn \\
&&\biggl.\left.-(\eta -\lambda +1) \left(-\lambda +q^2 (\eta +\lambda -2 \xi +1)-1\right)Y(r)\right)\biggr] = 8 \pi G_{\text{ae}} p_t(r)\,. \label{Y3}
\ea
Moreover the modified Einstein equation (1-2) can be written as

\be
Y'(r)=\frac{8-8 \lambda +W(r) \left[-4 \lambda +4 \xi +(\eta -\lambda +1) Y(r)\right]-2 Y(r) (\eta +\lambda -2 \xi +1)}{2 r (\eta -\lambda +1)}\,, \label{Y4}
\ee
\vskip 0.2cm
\noindent which has already been substituted in Eqs.~\eqref{Y1}--\eqref{Y3}.
Finally, Eq.~\eqref{hleq} for the scalar field and Eq.~\eqref{conserv} for the stress-energy tensor conservation (after Eqs.~\eqref{Y1}--\eqref{Y3} have been used) become, respectively,

\ba
&&\frac{(2 W(r)-3 Y(r)) \left[-8 \lambda +W(r) (-4 \lambda +4 \xi +(\eta -\lambda +1) Y(r))-2 Y(r) (\eta +\lambda -2 \xi +1)+8\right]}{r^2} \nn \\
&&-\frac{2 W'(r) \left[-4 \lambda +4 \xi +(\eta -\lambda +1) Y(r)\right]}{r}-\frac{2 Y'(r) \left[-4 (\eta -\xi +1)+3 (\eta -\lambda +1) W(r)-3 (\eta -\lambda +1) Y(r)\right]}{r} \nn \\
&&+4 (\eta -\lambda +1) Y''(r)=0\,, \nn \\ \label{Y5}
\ea
and

\ba
&&\frac{(W(r)-Y(r)) \left[8 (\lambda -1)+W(r) (4 (\lambda -\xi )+(-\eta +\lambda -1) Y(r))+2 Y(r) (\eta +\lambda -2 \xi +1)\right]}{r^2} \nn \\
&&+\frac{W'(r) \left[-4 \lambda +4 \xi +(\eta -\lambda +1) Y(r)\right]}{r}+\frac{Y'(r) \left[-4 (\eta -\xi +1)+3 (\eta -\lambda +1) W(r)-2 (\eta -\lambda +1) Y(r)\right]}{r} \nn \\
&&-2 (\eta -\lambda +1) Y''(r)=0\,. \nn \\ \label{Y6}
\ea
However, by substituting Eq.~\eqref{Y4} and its first derivative in Eqs.~\eqref{Y5} and \eqref{Y6}, these are identically satisfied.
So, we are finally left with only four independent field equations, {\it i.e.}, Eqs.~\eqref{Y1}--\eqref{Y4}. This means that, by assigning one of the two metric coefficients, $A(r)$ or $B(r)$, the other can be obtained by solving the differential Eq.~\eqref{Y4}, and the thermodynamical variables $\rho(r)$, $p_r(r)$, and $p_t(r)$ can be read from Eqs.~\eqref{Y1},~\eqref{Y2} and~\eqref{Y3}, respectively.
In this way we have obtained a family of infinite exact analytical solutions of the aforementioned system of equations if one of the two metric functions $A(r)$ or $B(r)$ is assigned.
Notice that spherically symmetric solutions in Ho\v rava gravity are identical to those of Einstein-\ae ther theory; therefore, our conclusions will hold for both theories all the same~\cite{Blas:2010hb}.

\section{Solution which reproduces the potential for a constant-density fluid sphere}
\label{Sec4}

We are now ready to work out the full system of field equations~\eqref{Y1}--\eqref{Y4}. Let us notice that Eq.~\eqref{Y4} can be generically solved for $W(r)$, which turns out to be

\be
W(r) = \frac{2 \left(4 (\lambda -1)+r (\eta -\lambda +1) Y'(r)+Y(r) (\eta +\lambda -2 \xi +1)\right)}{-4 \lambda +4 \xi +(\eta -\lambda +1) Y(r)}\,,
\ee
\vskip 0.2cm
\noindent and by making $A(r)$ and $B(r)$, explicit the latter becomes

\be \label{eqB}
\frac{r B'(r)}{B(r)}=-\frac{2 \left(r^2 (-\eta +\lambda -1) A'(r)^2+r A(r) \left(r (\eta -\lambda +1) A''(r)+2 (\eta -\xi +1) A'(r)\right)+4 (\lambda -1) A(r)^2\right)}{A(r) \left(r (-\eta +\lambda -1) A'(r)+4 A(r) (\lambda -\xi )\right)}\,.
\ee
\vskip 0.2cm
\noindent The equation above can be solved by assigning one of the two metric functions $A(r)$ or $B(r)$. Then the system of field equations is closed, and a family of infinite exact and analytical solutions can be found.

As an illustrative example let us choose the analytic form of $A(r)$ which reminds the Newtonian potential of a fluid sphere of constant density: {\it i.e.}, $A(r)= a + b r^2$, where $a$ and $b$ are arbitrary constants. Moreover, this choice also corresponds to the well-known Tolman IV solution in GR for an isotropic fluid~\cite{Tolman:1939jz}.
Substituting in Eq.~\eqref{eqB} the expression given above for $A(r)$, we obtain

\be
\frac{r B'(r)}{B(r)}=\frac{-4 a^2 (\lambda -1)+2 a b r^2 (-3 \eta -3 \lambda +2 \xi +1)-2 b^2 r^4 (\eta +3 \lambda -2 \xi -1)}{\left(a+b r^2\right) \left(2 a (\lambda -\xi )-b r^2 (\eta -3 \lambda +2 \xi +1)\right)}\,.
\ee
\vskip 0.2cm
\noindent This is a first-order ordinary differential equation that can be easily integrated to give

\be
B(r) = \frac{B_0 r^{\frac{2 (\lambda -1)}{\xi -\lambda }} \left[2 a (\lambda -\xi )-b r^2 (\eta -3 \lambda +2 \xi +1)\right]^{1+\frac{2 \eta }{\eta -3 \lambda +2 \xi +1}+\frac{\lambda -1}{\lambda -\xi }}}{\left(a+b r^2\right)^2}\,,
\ee
\vskip 0.2cm
\noindent where $B_0$ is an integration constant. 

It is now straightforward to get algebraically from Eqs.~\eqref{Y1}--\eqref{Y3} the explicit analytical expressions for the thermodynamical quantities $\rho(r)$, $p_r(r)$, and $p_t(r)$ that are shown below:

\ba
\rho(r)&=&\frac{1}{64 \pi B_0 \xi  G_{\mbox{\footnotesize \ae}} (\eta -\lambda +1)}\left(a+b r^2\right)^2 r^{\frac{2 (\xi -1)}{\lambda -\xi }} \left(2 a (\lambda -\xi )-b r^2 (\eta -3 \lambda +2 \xi +1)\right)^{-\frac{2 \eta }{\eta -3 \lambda +2 \xi +1}+\frac{1-\lambda }{\lambda -\xi }-1} \nn \\
&&\times \Biggl[-\frac{16 (\eta  \lambda -\lambda  \xi +\xi ) \left(2 a^2 (\lambda -1)+a b r^2 (3 \eta +3 \lambda -2 \xi -1)+b^2 r^4 (\eta +3 \lambda -2 \xi -1)\right)}{\left(a+b r^2\right) \left(2 a (\lambda -\xi )-b r^2 (\eta -3 \lambda +2 \xi +1)\right)}\Biggr. \nn \\
&&+\frac{8 B_0 \xi  (\eta -\lambda +1) r^{\frac{2 (\lambda -1)}{\xi -\lambda }} \left(2 a (\lambda -\xi )-b r^2 (\eta -3 \lambda +2 \xi +1)\right)^{\frac{2 \eta }{\eta -3 \lambda +2 \xi +1}+\frac{\lambda -1}{\lambda -\xi }+1}}{\left(a+b r^2\right)^2} \nn \\
&&\Biggl.-\frac{4 b r^2 \left(\lambda +q^2 (\eta -\lambda +1)-1\right) \left(4 a (\xi -\lambda )+b r^2 (\eta -5 \lambda +4 \xi +1)\right)}{\left(a+b r^2\right)^2}+8 \left(-\eta +2 \lambda ^2-3 \lambda\right. \nn \\
&&\Biggl.\left.+q^2 (\eta -\lambda +1) (2 \lambda -\xi -1)+1\right)\Biggr],
\ea
\ba
p_r(r)&=&\frac{1}{64 \pi B_0 \xi  G_{\mbox{\footnotesize \ae}}}\left(a+b r^2\right)^2 r^{\frac{2 (\xi -1)}{\lambda -\xi }} \left(2 a (\lambda -\xi )-b r^2 (\eta -3 \lambda +2 \xi +1)\right)^{-\frac{2 \eta }{\eta -3 \lambda +2 \xi +1}+\frac{1-\lambda }{\lambda -\xi }-1} \nn \\ 
&&\times \Biggl[\frac{4 b^2 r^4 \left(\lambda +q^2 (\eta -\lambda +1)-1\right)}{\left(a+b r^2\right)^2}-\frac{8 B_0 \xi  r^{\frac{2 (\lambda -1)}{\xi -\lambda }} \left(2 a (\lambda -\xi )-b r^2 (\eta -3 \lambda +2 \xi +1)\right)^{\frac{2 \eta }{\eta -3 \lambda +2 \xi +1}+\frac{\lambda -1}{\lambda -\xi }+1}}{\left(a+b r^2\right)^2} \Biggr. \nn \\
&&\Biggl.+\frac{16 b r^2 \left(\lambda -\lambda  q^2+\xi  q^2\right)}{a+b r^2}+16 \lambda +8 q^2 (-2 \lambda +\xi +1)-8\Biggr], \label{radialp}
\ea
and

\ba
p_t(r)&=&\frac{1}{64 \pi B_0 \xi  G_{\mbox{\footnotesize \ae}} (\eta -\lambda +1)}\left(a+b r^2\right)^2 r^{\frac{2 (\xi -1)}{\lambda -\xi }} \left(2 a (\lambda -\xi )-b r^2 (\eta -3 \lambda +2 \xi +1)\right)^{-\frac{2 \eta }{\eta -3 \lambda +2 \xi +1}+\frac{1-\lambda }{\lambda -\xi }-1} \nn \\
&&\times\Biggl[-8\Biggl(\frac{-2 \eta  \lambda +\eta +2 \lambda  \xi -3 \lambda +q^2 (\eta  (2 \lambda -\xi -1)-(\xi -1) (3 \lambda -2 \xi -1))+1}{\left(a+b r^2\right) \left(2 a (\lambda -\xi )-b r^2 (\eta -3 \lambda +2 \xi +1)\right)}\Biggr)\Biggr. \nn \\
&&\times \Biggl(2 a^2 (\lambda -1)+a b r^2 (3 \eta +3 \lambda -2 \xi -1)+b^2 r^4 (\eta +3 \lambda -2 \xi -1)\Biggr) \nn \\
&&+\frac{2 b r^2}{a+b r^2}\Biggl(-\frac{2 b r^2 (\eta -\lambda +1) \left(-\lambda +q^2 (\eta +\lambda -2 \xi +1)-1\right)}{a+b r^2}-4 \xi  \left(q^2 (\eta +3 \lambda +1)-2 \lambda \right)\Biggr. \nn \\
&&\Biggl.\Biggl.+4 \left(q^2-1\right) (\eta +\lambda  (2 \lambda -1)+1)+8 \xi ^2 q^2\Biggr)+16 (\lambda -1) \left(\lambda  \left(q^2-1\right)-\xi  q^2\right)\Biggr]. 
\ea

We also calculate the limit of the above expressions for large values of the radius $r\gg1$, when the constants $a$ and $b$ are sufficiently small:

\ba
\rho(r\gg1) &=& \frac{1}{64 \pi B_0 \xi G_{\text{ae}} (\eta -\lambda +1)}b^2 r^{\frac{2 (\xi -1)}{\lambda -\xi }+4} \left(-b r^2 (\eta -3 \lambda +2 \xi +1)\right)^{-\frac{2 \eta }{\eta -3 \lambda +2 \xi +1}+\frac{1-\lambda }{\lambda -\xi }-1} \nn \\
&&\times\Biggl[\frac{8 B_0 \xi  (\eta -\lambda +1) r^{\frac{2 (\lambda -1)}{\xi -\lambda }-4} \left(-b r^2 (\eta -3 \lambda +2 \xi +1)\right)^{\frac{2 \eta }{\eta -3 \lambda +2 \xi +1}+\frac{\lambda -1}{\lambda -\xi }+1}}{b^2} \Biggr. \nn \\
&&+\frac{16 (\eta +3 \lambda -2 \xi -1) (\eta  \lambda -\lambda  \xi +\xi )}{\eta -3 \lambda +2 \xi +1} +8 \left(-\eta +2 \lambda ^2-3 \lambda +q^2 (\eta -\lambda +1) (2 \lambda -\xi -1)+1\right) \nn \\
&&\Biggl. -4 (\eta -5 \lambda +4 \xi +1) \left(\lambda +q^2 (\eta -\lambda +1)-1\right)\Biggr],
\ea
\ba
p_r(r\gg1) &=& \frac{1}{64 \pi B_0 \xi  G_{\text{ae}}} b^2 r^{\frac{2 (\xi -1)}{\lambda -\xi }+4} \left(-b r^2 (\eta -3 \lambda +2 \xi +1)\right)^{-\frac{2 \eta }{\eta -3 \lambda +2 \xi +1}+\frac{1-\lambda }{\lambda -\xi }-1} \nn \\ 
&&\times\Biggl[-\frac{8 B_0 \xi  r^{\frac{2 (\lambda -1)}{\xi -\lambda }-4} \left(-b r^2 (\eta -3 \lambda +2 \xi +1)\right)^{\frac{2 \eta }{\eta -3 \lambda +2 \xi +1}+\frac{\lambda -1}{\lambda -\xi }+1}}{b^2}+16 \lambda +4 \left(\lambda +q^2 (\eta -\lambda +1)-1\right)\Biggr. \nn \\
&&\Biggl.+8 q^2 (-2 \lambda +\xi +1)+16 \left(\lambda -\lambda  q^2+\xi  q^2\right)-8\Biggr],
\ea
and 

\ba
p_t(r\gg1) &=& \frac{1}{16 \pi B_0 \xi  G_{\text{ae}} (\eta -3 \lambda +2 \xi +1)^2}
b r^{\frac{2 (\lambda -1)}{\lambda -\xi }}\left(-b r^2 (\eta -3 \lambda +2 \xi +1)\right)^{-\frac{2 \eta }{\eta -3 \lambda +2 \xi +1}+\frac{1-\lambda}{\lambda -\xi}}  \nn \\
&&\times \left[(3 \lambda -1) (\eta +9 \lambda -6 \xi -3)+q^2 \left(\eta ^2+\eta  (-6 \lambda +4 \xi +2)-3 (-3 \lambda +2 \xi +1)^2\right)\right].
\ea

The implementation of the junction conditions to the exterior vacuum metric amounts to requiring that the radial pressure $p_r(r)$ in Eq.~\eqref{radialp} vanish at $r=\bar{R}$. Then, $p_r[\bar{R}]=0$ results in

\ba
&&\frac{1}{64 \pi B_0 \xi  G_{\mbox{\footnotesize \ae}}}\left(a+b \bar{R}^2\right)^2 \bar{R}^{\frac{2 (\xi -1)}{\lambda -\xi }} \left(2 a (\lambda -\xi )-b \bar{R}^2 (\eta -3 \lambda +2 \xi +1)\right)^{-\frac{2 \eta }{\eta -3 \lambda +2 \xi +1}+\frac{1-\lambda }{\lambda -\xi }-1} \nn \\
&&\Biggl[\frac{4 b^2 \bar{R}^4 \left(\lambda +q^2 (\eta -\lambda +1)-1\right)-8 B_0 \xi  \bar{R}^{\frac{2 (\lambda -1)}{\xi -\lambda }} \left(2 a (\lambda -\xi )-b \bar{R}^2 (\eta -3 \lambda +2 \xi +1)\right)^{\frac{2 \eta }{\eta -3 \lambda +2 \xi +1}+\frac{\lambda -1}{\lambda -\xi }+1}}{\left(a+b \bar{R}^2\right)^2}\Biggr. \nn \\
&&\Biggl.+\frac{16 b \bar{R}^2 \left(\lambda -\lambda  q^2+\xi  q^2\right)}{a+b \bar{R}^2}+16 \lambda +8 q^2 (-2 \lambda +\xi +1)-8\Biggr]=0\,,
\ea
which is solved by

\be
q = \pm \Biggl[\frac{2 B_0 \xi  \bar{R}^{\frac{2 (\lambda -1)}{\xi -\lambda }} \left(2 a (\lambda -\xi )-b \bar{R}^2 (\eta -3 \lambda +2 \xi +1)\right)^{\frac{2 \eta }{\eta -3 \lambda +2 \xi +1}+\frac{\lambda -1}{\lambda -\xi }+1}-b \bar{R}^2 (3 \lambda -1) \left(3 b \bar{R}^2+4 a\right)+2 a^2 (1-2 \lambda )}{4 a b \bar{R}^2 (-3 \lambda +2 \xi +1)+b^2 \bar{R}^4 (\eta -9 \lambda +6 \xi +3)+2 a^2 (-2 \lambda +\xi +1)}\Biggr]^{\frac{1}{2}}.
\ee
\vskip 0.2cm
At this stage, we set $G_{\mbox{\footnotesize \ae}}=G_N \left(1-\eta/2\xi \right)$, where $G_N$ is the Newton's constant, which is needed to recover the Newtonian limit~\cite{Carroll:2004ai,Blas:2009qj}. Moreover, we also implement the constraint coming from the near-simultaneous observation of the gravitational-wave event GW170817 and the gamma-ray burst GRB 170817A~\cite{Monitor:2017mdv}, which consists in setting the speed of propagation of the spin-2 mode to 1, {\it i.e.}, $\xi=1$, up to an uncertainty of about $10^{-15}$~\cite{Monitor:2017mdv,Gumrukcuoglu:2017ijh}.

The outcome of such analysis is that the solution worked out above cannot be considered valid in the whole interior spacetime. Indeed, it is plagued by a singularity in the center as the curvature invariants $R$, $R_{\mu\nu}R^{\mu\nu}$, and $R_{\alpha\beta\mu\nu}R^{\alpha\beta\mu\nu}$ diverge at $r=0$; then the corresponding spacetime is also not geodetically complete because of the singularity that it inherits. For the sake of simplicity, we only write below the corresponding expression for the scalar curvature $R$, which is
\ba
R&=&-\frac{2}{B_0 r^2} \left(2 a (\lambda -1)-b r^2 (\eta -3 \lambda +3)\right)^{-\frac{2 \eta }{\eta -3 \lambda +3}-4}\Biggl[-2 b r^4 \left(b r^2 (\eta -3 \lambda +3)-2 a (\lambda -1)\right)\Biggr. \nn \\
&&\left(-4 a^2 (\lambda -1)-9 a b (\lambda -1) r^2+b^2 r^4 (\eta -6 \lambda +6)\right)+2 r^2 \left(a+b r^2\right) \left(b r^2 (\eta -3 \lambda +3)-2 a (\lambda -1)\right) \nn \\
&&\left(2 a^2 (\lambda -1)+3 a b r^2 (\eta +\lambda -1)+b^2 r^4 (\eta +3 \lambda -3)\right)+b^2 r^6 \left(b r^2 (\eta -3 \lambda +3)-2 a (\lambda -1)\right)^2 \nn \\
&&\Biggl.+B_0 \left(2 a (\lambda -1)-b r^2 (\eta -3 \lambda +3)\right)^{\frac{2 \eta }{\eta -3 \lambda +3}+4}-r^2 \left(a+b r^2\right)^2 \left(b r^2 (\eta -3 \lambda +3)-2 a (\lambda -1)\right)^2\Biggr].
\ea
It is easy to show that the expression above diverges in the center as $\sim 1/r^2$. Moreover, the thermodynamical quantities $\rho(r)$ and $p_r(r)$ take infinite values at $r=0$. 
Then, the internal spacetime described by this solution loses its physical predictability at very small scales in the interior of astrophysical objects.
Nevertheless, one might still consider this kind of solution as a viable model in the context of a star with several internal shells. Indeed, in that case, the solution at hand would only hold from the surface towards a certain internal physical radius, while from the latter up to to the center of the fluid sphere a new interior metric would be needed. 

Anyhow, this proof shows that in the context of Ho\v rava gravity and Einstein-\ae ther theory with a non-static \ae ther of the form given by Eq.~\eqref{aether2}, it is not possible to construct a viable solution whose metric coefficient $A(r)$ reproduces the potential of a Newtonian constant density sphere across the whole interior spherically symmetric spacetime. Similar results can be obtained for several choices of $A(r)$, which means that the specific choice considered here does not imply a loss of generality, since qualitatively the conclusions do not change by making a different ansatz. Notice that if the \ae ther is assumed to be static instead, such a solution exists~\cite{Vernieri:2017dvi,Vernieri:2018sxd}. One possibility to resolve this issue might be to relax the hypothesis made for the function $F(r)$ and then to consider even more general configurations of the \ae ther vector field.

\section{Conclusions}
\label{Sec5}

We have taken into account spherically symmetric interior spacetimes filled by anisotropic fluids in the context of Ho\v rava gravity and Einstein-\ae ther theory. A general setting in which the \ae ther vector field is non-static has been implemented, which means that the \ae ther has two non-trivial components, instead of a single one as in the case of a static \ae ther. We have anyhow made some assumption on the component $F(r)$ of the \ae ther by means of which the field equations become analytically solvable. Then, we have shown that a family of infinite exact analytical solutions exists when one of the two metric coefficients is assigned. The result is quite remarkable, since these are the first exact and analytical solutions ever found in the context of such theories in presence of anisotropic matter and with a non-static \ae ther. As an illustrative example, we have solved the field equations by selecting the metric coefficient $A(r)$ in such a way as to reproduce the Newtonian potential for a fluid sphere of constant density, which also coincides with the Tolman IV solution in GR for an isotropic fluid.
Nevertheless, the resulting analytical solution that we have found for the metric coefficients and the thermodynamical quantities is plagued by an unavoidable divergence in the center. Indeed, all the curvature invariants are found to diverge at $r=0$. For simplicity, we have displayed only the expression of the scalar curvature $R$ which diverges with a power of $r^{-2}$. 

In the framework of a quantum gravity theory like Ho\v rava gravity, that should also account for strong-gravity effects, the presence of any kind of singularity in the description of astrophysical objects is somehow questionable. Most likely, this fact might signal that such a solution is not admissible only in the context that we have considered here. 
Notice that, both in the case of static \ae ther and isotropic fluids~\cite{Eling:2006df,Eling:2007xh}, and in the case of static \ae ther and anisotropic fluids~\cite{Vernieri:2017dvi,Vernieri:2018sxd}, such a singularity for spherically symmetric interior spacetimes is not present. As a direct consequence of this, one can infer that the main physical difference between those scenarios and the one under study in this paper is due to the choice of a non-static \ae ther that we are performing here, and not due to the anisotropy of pressure, which indeed does not seem to play any relevant role in this respect.
So, further investigations are necessary in order to understand if more general configurations of the \ae ther vector field can solve this issue. Moreover, an alternative route could be to consider in the gravitational action also the higher-order operators which render the theory power-counting renormalizable. Those extra operators might indeed cure the pathology which plagues the solution in the center. In both cases, one would need to use numerical approaches in order to solve the resulting field equations. 

\vspace{0.3cm}
{\bf Acknowledgments:} The author would like to thank Sante Carloni for enlightening discussions that led to the initiation of this project. This research was supported by Funda\c{c}\~ao para a Ci\^encia e a Tecnologia (FCT) through the Research Grant UID/FIS/04434/2019, and by Projects PTDC/FIS-OUT/29048/2017, COMPETE2020: POCI-01-0145-FEDER-028987 $\&$ FCT: PTDC/FIS-AST/28987/2017, and IF/00852/2015 of FCT.


\end{document}